\documentclass[10pt]{iopart}
\usepackage{iopams,graphicx}

\begin{document}

\title
{Binary Reactive Adsorbate on a Random Catalytic Substrate}

\author{
M N Popescu\dag\footnote[3]{To whom correspondence should 
be addressed}, S Dietrich\dag~and G Oshanin\ddag\dag}

\address{\dag\ Max-Planck-Institut f{\"u}r Metallforschung,
Heisenbergstr. 3, 70569 Stuttgart, Germany, and\\
Institut f{\"u}r Theoretische und Angewandte Physik,
Universit{\"a}t Stuttgart, Pfaffenwaldring 57, 70569 Stuttgart,
Germany}

\address{\ddag\ Laboratoire de Physique Th{\'e}orique de la 
Mati{\`e}re Condens{\'e}e,
Universit{\'e} Paris 6, 4 Place Jussieu, 75252 Paris, France}

\eads{\mailto{popescu@mf.mpg.de}, \mailto{dietrich@mf.mpg.de},
\mailto{oshanin@lptl.jussieu.fr}}

\begin{abstract}

We study the equilibrium properties of a model for a 
binary mixture of catalytically-reactive monomers adsorbed 
on a two-dimensional substrate decorated by randomly 
placed catalytic bonds. The interacting $A$ and $B$ monomer 
species undergo continuous exchanges with particle reservoirs 
and react ($A + B \to \emptyset$) as soon as a pair of unlike 
particles appears on sites connected by a catalytic bond. 
For the case of annealed disorder in the placement of the 
catalytic bonds this model can be mapped onto a classical 
spin model with spin values $S = -1,0,+1$, with effective 
couplings dependent on the temperature and on the mean 
density $q$ of catalytic bonds. This allows us to exploit the 
mean-field theory developed for the latter to determine the 
phase diagram as a function of $q$ in the (symmetric) case 
in which the chemical potentials of the particle reservoirs, 
as well as the $A-A$ and $B-B$ interactions are equal.

\end{abstract}




\section{Introduction}
\label{intro}
Catalytically activated reactions (CARs) involve particles 
which react only in the presence of another agent acting as 
a catalyst, and remain chemically inactive otherwise. 
Usually, the catalyst is part of a solid, inert substrate 
placed in contact with fluid phases of the reactants, and 
the reaction takes place only between particles adsorbed 
on the substrate forming a (dilute) monolayer. These processes 
are widespread in nature and used in a variety of technological 
and industrial applications \cite{1b}.\nolinebreak

The work of Ziff, Gulari, and Barshad (ZGB) \cite{zgb} on 
the ``monomer-dimer'' model, introduced as an idealized 
description of the process of ${\rm CO}$ oxidation on a catalytic 
surface, as well as the subsequent studies of a simpler 
``monomer-monomer'' reaction model \cite{fich_red}, represent 
an important step in the understanding of CARs properties by 
revealing the emergence of an essentially collective behavior in 
the dynamics of the adsorbed monolayer. On two-dimensional (2D) 
substrates, first- and second-order non-equilibrium phase 
transitions involving saturated, inactive phases (substrate 
poisoning, i.e., most of the adsorption sites are occupied by 
same-type particles) and reactive steady-states have been 
evidenced and studied in detail 
\cite{zgb,fich_red,con,universality_class,dic}. Most of these 
available studies pertain to idealized homogeneous substrates.

In contrast, the equilibrium properties of the adsorbed monolayer 
in the case of CARs are much less studied and the understanding 
of the equilibrium state remains rather limited. Moreover, 
actual substrates are typically disordered and generically the  
catalyst is an assembly of mobile or localized catalytic sites 
or islands \cite{1b}; the recently developed artificially designed 
catalysts \cite{Abbet_00} involve inert substrates which are 
decorated by catalytic particles. Theoretical studies which 
have addressed the behavior of CARs on disordered substrates 
have been so far focused on the effect of site-dependent 
adsorption/desorption rates because natural catalysts are, 
in general, energetically heterogeneous \cite{reviews,dis}; 
only few studies, in particular some exactly solvable 1D models 
of $A + A \to \emptyset$ reactions and a Smoluchowski-type 
analysis of $d$-dimensional CARs \cite{gleb}, have addressed 
the case of spatially heterogeneous catalyst distribution.

Recently, we have presented a simple model of a monomer-monomer  
$A + B \to \emptyset$ reaction on a 2D inhomogeneous, catalyst 
decorated substrate, and we have shown that for the case of 
\textit{annealed} disorder in the placement of the catalytic bonds 
the reaction model under study can be mapped onto the general 
spin $S = 1$ (GS1) model \cite{berker} with effective, temperature 
dependent couplings \cite{opd_04}. This allows us to exploit the 
large number of results obtained for the GS1 model 
\cite{berker,mean_field} in order to elucidate, within a mean-field 
description \cite{mean_field}, the equilibrium properties of the 
monolayer binary-mixture of reactive monomers on a 2D substrate 
randomly-decorated by a catalyst.

The organization of the paper is as follows. In 
Sec.~\ref{model} we briefly present the model for a 
binary-mixture monolayer with an 
$A + B \stackrel{catalyst}{\longrightarrow}\emptyset$ 
reaction on a 2D inhomogeneous, catalyst decorated substrate 
and the mapping to a GS1 model; in Sec.~\ref{mean_field} we 
present the mean-field (MF) approximation. 
Section~\ref{q0_and_q1} is devoted to a discussion of the MF 
phase diagram for the particular cases of a completely 
catalytic substrate, i.e., $q = 1$, and of an inert substrate, 
i.e., $q = 0$, respectively. In Sec.~\ref{q_gen} we discuss, 
on the basis of the results for $q = 0$, the MF phase diagram 
for general values of $q$. We conclude with a brief summary 
of the results in Sec.~\ref{summary}.

\section{Model of a monolayer binary mixture of reactive 
species on a 2D inhomogeneous, catalyst decorated substrate}
\label{model}
\begin{figure}[!htb]
\begin{center}
\includegraphics[width=.55 \linewidth]{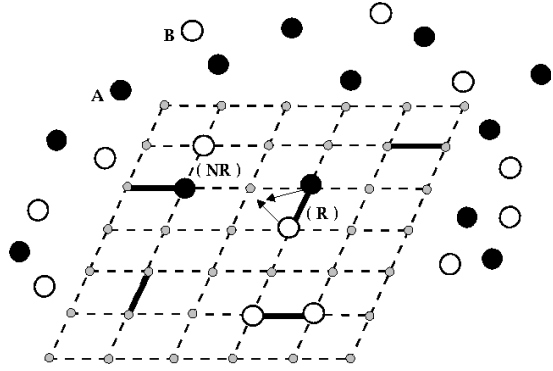}
\caption{
\label{fig1}
2D lattice of adsorption sites (small grey circles) in contact 
with a mixed vapor phase. Black (white) circles denote $A$ ($B$) 
particles, respectively, solid lines denote ``catalytic bonds''. 
(\textbf{R}): configuration in which an annihilation reaction 
({\tiny $\nearrow \nwarrow$}) takes place;
(\textbf{NR}): NN pair $A$-$B$, but no reaction because there 
is no catalytic bond between these sites.
}
\end{center}
\end{figure}

We consider a 2D regular lattice (coordination number $z$) of 
$N$ adsorption sites (Fig.~\ref{fig1}), which is in contact 
with the mixed vapor phase of $A$ and $B$ particles.
The $A$ and $B$ particles can adsorb onto $vacant$ sites, and can 
desorb back to the reservoir. The system is characterized by 
chemical potentials $\mu_{A,B}$ maintained at constant values and 
measured relative to the binding energy of an occupied site, so 
that $\mu_{A,B} > 0$ corresponds to a preference for adsorption. 
Both $A$ and $B$ particles have hard cores prohibiting double 
occupancy of the adsorption sites and nearest-neighbor (NN) 
attractive $A-A$, $B-B$, and $A-B$ interactions of strengths 
$J_A$, $J_B$, and $J_{AB}$, respectively. The occupation of the 
$i$-th site is described by a ``spin'' variable
\begin{equation}
\sigma_i =
\cases{
+(-)1, & ~~\textrm{site $i$ occupied by $A$ ($B$),}\\
~0,  & ~~\textrm{site $i$ empty.}\\
}
\label{spin}
\end{equation}
We assign, at random, to some of the lattice bonds (solid lines 
in Fig.~\ref{fig1}) ``catalytic'' properties such that 
if an $A$ and a $B$ particle occupy simultaneously NN sites 
connected by such a catalytic bond, they instantaneously react 
and desorb, and the product ($AB$) leaves the system; $A$ and $B$ 
particles occupying NN sites not connected by a 
catalytic bond harmlessly coexist, and we assume that the reverse 
process of a simultaneous adsorption of an $A$ and a $B$ on a 
catalytic bond has an extremely low probability and can be neglected.
The ``catalytic'' character of the lattice bonds is described by 
variables $\zeta_{<ij>}$, where $<ij>$ denotes a pair of NN sites 
$i$ and $j$, 
\begin{equation}
\zeta_{<ij>} =
\cases{
1, & ~~\textrm{$<ij>$~is a catalytic bond,} \\
0,  & ~~\textrm{otherwise,}\\}
\label{zeta}
\end{equation}
and we take $\{\zeta_{<ij>}\}$ as independent, identically 
distributed random variables with the probability distribution
\begin{equation}
\varrho(\zeta) =  q \delta(\zeta - 1) + (1-q)\delta(\zeta).
\label{dist_zeta}
\end{equation}
Note that the probability $q$ that a given bond is catalytic 
equals the mean density of the catalytic bonds. The two limiting 
cases, $q = 0$ and $q = 1$, correspond to an \textit{inert} 
substrate and to a \textit{homogeneous catalytic} one, 
respectively. We further assume that the condition of instantaneous 
reaction $A + B \stackrel{catalyst}{\longrightarrow}\emptyset$ 
together with negligible simultaneous adsorption of an $A$ 
and a $B$ particle on a catalytic bond is formally 
equivalent to allowing a NN $A-B$ repulsive interaction of 
strength $\lambda \gg 1$, followed by the limit 
$\lambda \to \infty$, for $A-B$ pairs connected by catalytic 
bonds.

As shown in Ref.\cite{opd_04}, in thermal equilibrium and for 
situations in which the disorder in the placement of the 
catalytic bonds is \textit{annealed}, i.e., the partition 
function, rather than its logarithm, is averaged over the 
disorder, the model under study is mapped exactly onto that 
of a GS1 model. The "effective" GS1 Hamiltonian describing the 
adsorbate at temperature $T$ is:
\begin{equation}
\fl\hspace*{.5cm}\mathcal{H}_{e} = - J\sum_{<ij>}\sigma_i \sigma_j 
- K \sum_{<ij>}\sigma_i^2 \sigma_j^2 
-C\sum_{<ij>}\left(\sigma_i \sigma_j^2+\sigma_j \sigma_i^2\right)
-H \sum_{i=1}^{N} \sigma_i + \Delta \sum_{i=1}^{N} \sigma_i^2\,,
\label{H_BEG}
\end{equation}
where the coupling constants are given explicitely by:
\begin{eqnarray}
\label{param}
J = \frac{J_A+J_B - 2 J_{AB}}{4} - \frac{k_B T}{2} \ln(1-q)
:= J_0 - \frac{k_B T}{2} \ln(1-q) \,, \nonumber\\
K = \frac{J_A+J_B + 2 J_{AB}}{4} + \frac{k_B T}{2} \ln(1-q)
:= K_0 + \frac{k_B T}{2} \ln(1-q) \,,\\
C = \frac{J_A-J_B}{4},\; H = \frac{\mu_A -\mu_B}{2},\;
\Delta = -\frac{\mu_A +\mu_B}{2},\nonumber
\end{eqnarray}
and $T$ is the temperature. In the remaining part of this paper 
we focus on the symmetric case in which the chemical potentials of 
the two species are equal, $\mu_A = \mu_B := \mu$, (implying $H = 0$), 
and $J_A = J_B := j$, (implying $C = 0$). This model reduces to the 
original Blume-Emery-Griffiths (BEG) model \cite{mean_field} in zero 
magnetic field $H$.

\section{Mean-field approximation of the free energy.}
\label{mean_field}
The mean-field analysis follows closely the presentation in 
Ref.\cite{mean_field} and thus here we only briefly outline 
the main steps. The starting point is the variational principle 
for the free energy $F$ (see, e.g., Ref.\cite{variational_f})
\begin{equation}
F \leq \Phi[\rho] := 
\Tr(\rho \mathcal{H}_{e}) + k_B T \Tr(\rho \ln \rho)\,
\label{variational}
\end{equation}
where $\rho$ is any trial density matrix, i.e., $\Tr(\rho) = 1$; 
the equality holds for $\rho = 
\exp(-\beta \mathcal{H}_{e})/\Tr[\exp(-\beta \mathcal{H}_{e})]$, 
where $\beta^{-1} = k_B T$ and $\Tr$ denotes the sum over all 
spin configurations. 

Within the mean-field approximation the trial density $\rho$ is 
chosen from the subspace of products of single-site densities, 
i.e., $\rho = \prod\limits_{i=1}^{N}\rho_i$; 
furthermore restricting to the case of translationally invariant 
states, i.e., $\rho_i$ being independent of $i$, leads to trial 
densities of the form $\rho = \rho_1^N$. Note that this last 
restriction implies that within the present approximation the 
analysis cannot account for the occurrence of staggered states 
(i.e., splitting in ordered sub-lattices) and the emphasis is put on 
the disordered and ordered homogeneous states. The single site 
density $\bar\rho_1$ minimizing the functional $\Phi/N$, subject 
to the constraint $\Tr(\bar\rho_1) = 1$, is 
\begin{eqnarray}
\bar \rho_1 &= \exp(-\beta h)/\Tr[\exp(-\beta h)]\,,\nonumber\\
h &= -J' M \sigma_1 + (\Delta - K' Q) \sigma_1^2\,,
\label{rho_min}
\end{eqnarray}
where $J' = z J$, $K'= z K$, and 
\begin{eqnarray}
M := 
\langle \sigma_1 \rangle \equiv \Tr(\bar\rho_1 \sigma_1)\,,\nonumber\\
Q := 
\langle \sigma_1^2 \rangle \equiv \Tr(\bar\rho_1 \sigma_1^2)\,
\label{M_and_Q}
\end{eqnarray}
are the so-called magnetization $M$ and the quadrupolar moment 
$Q$, respectively. This leads to the following approximation 
$f_{mf}$ of the free energy per-site:
\begin{equation}
\fl f_{mf}(M,Q) = \Phi[\bar\rho_1]/N = 
\frac{1}{2}(J' M^2 + K' Q^2) - \frac{1}{\beta} 
\ln\left[1+ 2 e^{-\beta \Delta} e^{\beta K' Q}\cosh(\beta J' M)\right]\,.
\label{f_minimize}
\end{equation}
Note that in the binary mixture language the values of the 
magnetization and of the quadrupolar moment [Eq.(\ref{M_and_Q})] 
represent the difference and the sum (total coverage) of 
the average densities $n_A$ and $n_B$ of $A$ and $B$ species, 
respectively: 
\begin{equation}
M = n_A - n_B \,,\,\,\,Q = n_A + n_B \,.
\label{eq_densities}
\end{equation}

For given values of the temperature $T$ and of the field 
$\Delta = -\mu$ [Eq.(\ref{param})], the order parameters $M$ and 
$Q$ are obtained by solving Eqs.(\ref{rho_min}) and 
(\ref{M_and_Q}). The pair $(M,Q)$ characterizing the state of the 
system is selected from the possible solutions as the one which 
minimizes $f_{mf}$ in Eq.(\ref{f_minimize}) above. Explicitely, the 
equations determining $M$ and $Q$ are:
\begin{equation}
M = 
\frac{2 \sinh(\beta J' M)}{\exp(-\beta \mu - \beta K'Q)+ 
2 \cosh(\beta J' M)}\,,
\label{eq_M}
\end{equation}
\begin{equation}
Q =
\frac{2 \cosh(\beta J' M)}{\exp(-\beta \mu - \beta K'Q)+ 
2 \cosh(\beta J' M)}\,.
\label{eq_Q}
\end{equation}
Note that there is always a solution of these equations 
with $M = 0$, i.e., a disordered state (or, in the language 
of binary mixtures, a mixed state).

Alternatively, one may search directly for the absolute minimum 
of the per-site free-energy function with respect to $M$ and $Q$.
The values $M$ and $Q$ at the minimum (minima in case of phase 
equilibria) will define the thermodynamically stable phase(s). 
While the first formulation is more useful for analytical work, 
reduced to analyzing the number of solutions of two coupled 
algebraic equations, the latter is advantageous for numerical 
calculations. In the following we shall use both of them. 
Before proceeding we note that $f_{mf}(M,Q)$ [Eq. (\ref{f_minimize})] 
is an even function of $M$ (i.e., invariant under the change 
$M \to -M$), and thus in the following we shall restrict the 
discussion to the case $M \geq 0$; the states with $M \leq 0$ are 
immediately obtained via a change of sign. This is a consequence 
of the symmetry in the chemical potentials ($\mu_A = \mu_B$) or, 
in the magnetic language, of a vanishing magnetic field $H = 0$. 
In other words, in the space spanned by $(T,\mu, H)$ the phase 
diagrams in the plane $H = 0$ along the $H = 0^+$ side and any 
equilibrium state characterized by $(M > 0,Q)$ have corresponding 
phase diagrams and states $(-M,Q)$ located on the $H = 0^-$ side.

\section{Homogeneous catalytic or catalytically inert substrates.}
\label{q0_and_q1}
\subsection{The case of a homogeneous catalytic substrate: $q = 1$.}
\label{q_1}
The case of a homogeneous, completely catalytic substrate 
can be studied analytically because of the particular form 
of the interaction parameters $K'$ and $J'$. In the limit 
$q \to 1$, Eq. (\ref{param}) implies:
\begin{eqnarray}
J' = z J_0 - z \frac{k_B T}{2} \ln(1-q) 
\stackrel{q \to 1}{\longrightarrow} +\infty \,,\nonumber\\
K'  = z J_0 + z \frac{k_B T}{2} \ln(1-q) 
\stackrel{q \to 1}{\longrightarrow} -\infty \,,
\end{eqnarray}
while $J' + K' = z j$ remains finite in this limit. The analysis 
of Eqs.(\ref{eq_M}) and  (\ref{eq_Q}) proceeds as follows. 
From Eq.(\ref{eq_Q}), the solutions with $M = 0$, i.e., 
disorder states, have the quadrupolar moment
\begin{equation}
Q = \lim_{K'\to -\infty} 
\frac{2}{\exp(-\beta \mu) \exp(- \beta K' Q) + 2} = 0\,,
\label{Qdis_q1}
\end{equation}
for any finite temperature and finite chemical potential 
$\mu$. This corresponds to an empty lattice state.

Taking the ratio of Eqs.(\ref{eq_M}) and  (\ref{eq_Q}) we 
find that the ordered states, $M \neq 0$, satisfy
\begin{equation}
Q = \lim_{J'\to +\infty} M\coth(\beta J' M) = |M|\,.
\label{Qord_q1}
\end{equation}
In this limit the substrate is occupied by a single species, 
either $A$ or $B$ with equal probability. With $K' + J' = jz$, 
for any finite temperature and finite chemical potential 
$M$ is determined from 
\begin{eqnarray}
M = 
\frac{1-\exp(-2 \beta J' M)}
{\exp(\beta \Delta)\exp(-\beta j z M) + 1 + \exp(-2 \beta J' M)}
\nonumber\\ 
\stackrel{J'\to +\infty}{\longrightarrow}
\frac{1}{\exp(-\beta \mu)\exp(- \beta j z M) + 1}\,.
\label{Mord_q1}
\end{eqnarray}
It is easy to see that Eq.(\ref{Mord_q1}) has a solution 
$0 \leq |M| \leq 1$ for any $T > 0$ and any $\mu$. Thus, in 
virtue of Eq.(\ref{Qord_q1}), the lattice is occupied, with 
equal probability, either by $|M| \times N$ particles of 
species $A$ or by $|M|\times N$ particles of species $B$.

\subsection{The case of an inert substrate: $q = 0$.}
\label{q_0}
In the case of a catalytically inert substrate, $q = 0$, 
we have 
\begin{equation}
J'= z \frac{j - 2 J_{AB}}{2} = z J_0\,\,,\,\,
K'  = z \frac{j + 2 J_{AB}}{2} = z K_0 \,,
\label{J_K_q0}
\end{equation}
and thus the model reduces to the classical BEG model, whose 
mean-field approximation has been analyzed in detail in Ref. 
\cite{mean_field}. In the following we briefly summarize the 
main aspects of the phase diagram of the BEG model such that 
in the general case $q \neq 0$ we can isolate the effects 
solely due to the disordered distribution of the catalyst. 
Moreover, as will be shown in the next section, the equilibrium 
properties of the adsorbate in the general case $q \neq 0$ can 
be easily rationalized from the ones on the inert substrate.

First we note that for non-interacting particles, i.e., 
for $j = J_{AB} = 0$ so that $J' = K' = 0$, Eq.(\ref{eq_M}) 
implies that $M = 0$, and thus Eq.(\ref{eq_Q}) leads to
\begin{equation}
Q = \frac{2}{\exp(-\beta \mu)+ 2}\,\,,\,
n_A = n_B = Q/2 = \frac{\mathfrak{f}}{1 + 2 \mathfrak{f}}\,,
\label{eq_Q_Lang}
\end{equation}
where $\mathfrak{f} := \exp(\beta \mu)$ is the fugacity, 
rendering the classical Langmuir adsorption result.

For the case of non-zero interaction parameters $j$ and 
$J_{AB}$, only the qualitative features of the phase-diagram 
can be analytically derived (for details see Ref.\cite{mean_field}); 
here we shall present phase diagrams obtained via direct numerical 
minimization of the free energy function $f_{mf}$ 
[Eq.(\ref{f_minimize})]. Using $z J_0$ as energy scale, the 
system is characterized by the parameter $\kappa_0 := K'/J' = K_0/J_0$,  
the scaled temperature (thermal energy) $t$, and the scaled chemical 
potential $u$: 
\begin{equation}
t:= k_B T/(z J_0)\,\,,\,u:= \mu /(z J_0)\,\,.
\end{equation}
We shall discuss below the phase diagrams, as well as the 
behavior of the order parameters $M$ and $Q$, in the $u - t$ plane 
at given values of $\kappa_0 \geq 0$. The reason for the latter 
restriction is that for sufficiently negative values 
$\kappa_0 < \kappa_0^{(tr)} <0$, where $\kappa_0^{(tr)}$ is a 
threshold value, it is known that the system will split into 
ordered sub-lattices \cite{stagerred}. However, such states are 
not captured by the present formulation of the mean-field 
equations which assume translational invariance.

Before proceeding with the numerical analysis, we list several 
general features of the phase diagram that can be obtained from 
the analysis of Eqs.(\ref{eq_M}) and (\ref{eq_Q}).\hfill\\
\textbf{(i)} For $M \neq 0$, Eq.(\ref{eq_M}) can be written as 
\begin{equation}
\underbrace{2 [t x \cosh(x) - \sinh(x)]}_{\hspace*{1.cm}:=g_1(x;t)} 
+ \underbrace{t x \exp(-u/t) 
\exp(-\kappa_0 Q/t)}_{\hspace*{1.3cm}:=g_2(x;t,u)} 
= 0\,,
\label{eq_t}
\end{equation}
where $0 < x := M/t \leq 1/t$. Since $g_1(x;t>1)$ is a strictly 
increasing function of $x$, it follows that 
$g_1(x>0;t>1) > g_1(0;t>1) = 0$; moreover, one has 
$g_2(x>0;t>0,u) > 0$, from which one can conclude that for $t > 1$ 
Eq.(\ref{eq_t}) has only the solution $x = 0$, i.e., there are no ordered 
states for $t > 1$. \hfill\\
\textbf{(ii)} For $\mu < \mu_c(T) < 0$, the positive term 
$g_2(x;t,u < 0)$ dominates over the term $g_1(x;t)$, which is bounded 
from below, and thus in this range Eq. (\ref{eq_t}) has only the 
solution $x = 0$, i.e., there are no ordered states for 
$u < u_c(t) < 0$. 
This is an intuitive result: $\mu < 0$ corresponds to a 
preference for desorption; thus at low negative chemical potential 
the substrate is covered by a low density two-dimensional gas 
(which is a mixed phase).\hfill\\
\textbf{(iii)} For $\mu \to \infty$ and finite temperatures, 
$g_2(x;t,u \to \infty) \to 0$, and thus Eq.(\ref{eq_t}) reduces 
to $g_1(x;t) = 0$ which is equivalent to $\tanh(x) = t x$. 
It follows that for $t <1$ there is always a unique solution 
$x(t) > 0$, and therefore there is an ordered state 
$M(t < 1,\mu \to \infty) \neq 0$ such that 
$M(t \nearrow 1,\mu \to \infty) \to 0$, i.e., if there is a 
phase transition at $t = 1$ (for $\mu \to \infty$), it is a 
continuous order-disorder transition.\hfill\\
\textbf{(iv)} In the case of disordered states (i.e., $M=0$) 
Eq.(\ref{eq_Q}) reduces to 
\begin{equation}
\underbrace{2 t y + t y \exp(-u/t) \exp(-\kappa_0 y) 
- 2}_{\hspace*{1.2cm}:=g_3(y;u,t)} = 0\,,
\label{eq_u}
\end{equation}
where $0 < y := Q/t \leq 1/t$. For $\mu > 0$ and finite 
temperatures, $g_3(y;u>0,t)$ is a strictly increasing 
function of $y$, and thus Eq. (\ref{eq_u}) has an unique 
solution, i.e., in the region $\mu > 0$ there are no phase 
transitions between disordered states.

\begin{figure}[!htb]
\begin{center}
\centerline{\hspace*{.2\textwidth}(a) 
[M]\hspace*{.37\textwidth}(b) [Q]\hfill}
\begin{minipage}[c]{.45\textwidth}
\includegraphics[width = 0.9\textwidth,height=0.9\textwidth]{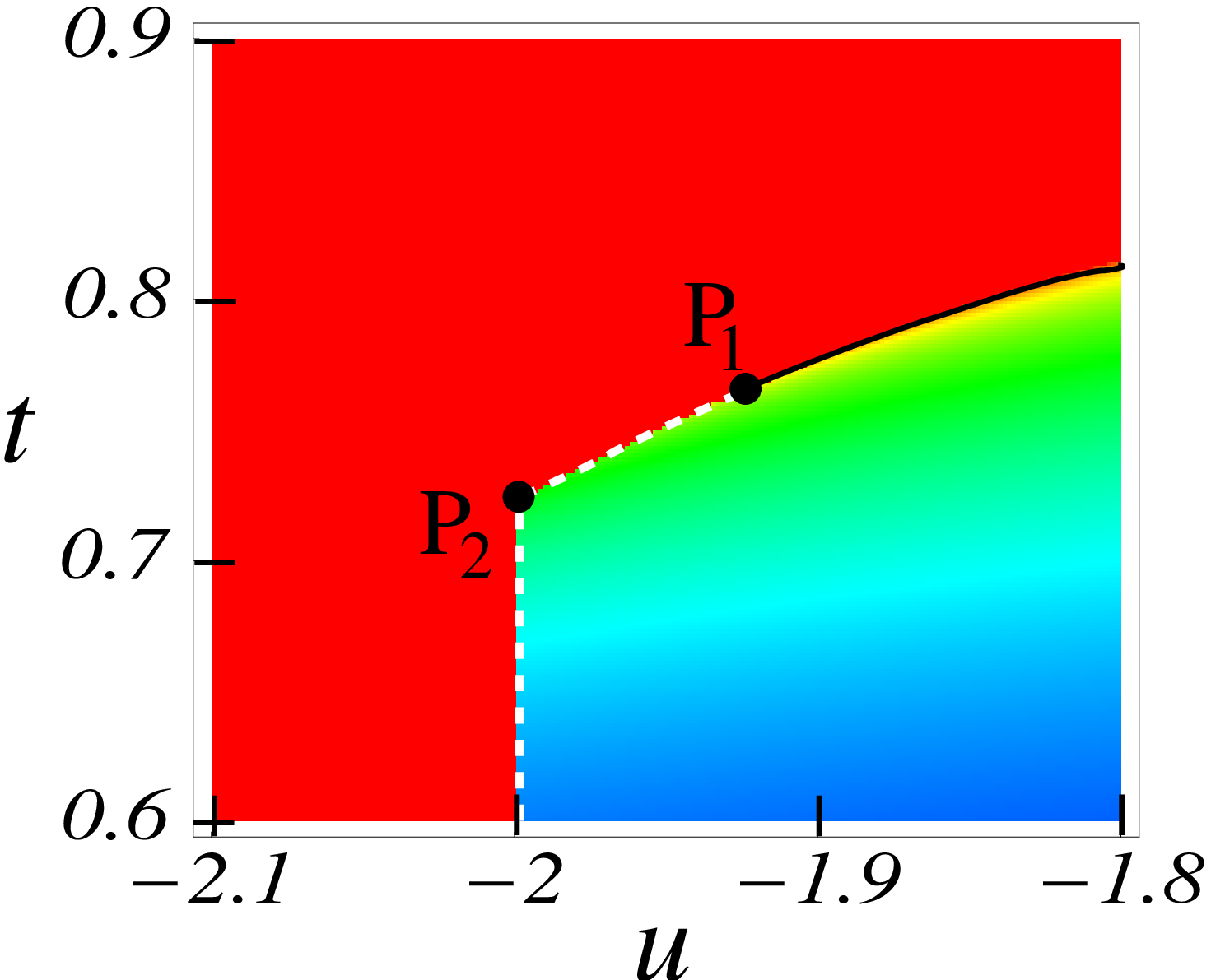}
\end{minipage}%
\begin{minipage}[c]{.45\textwidth}
\includegraphics[width = 0.9\textwidth,height=0.9\textwidth]{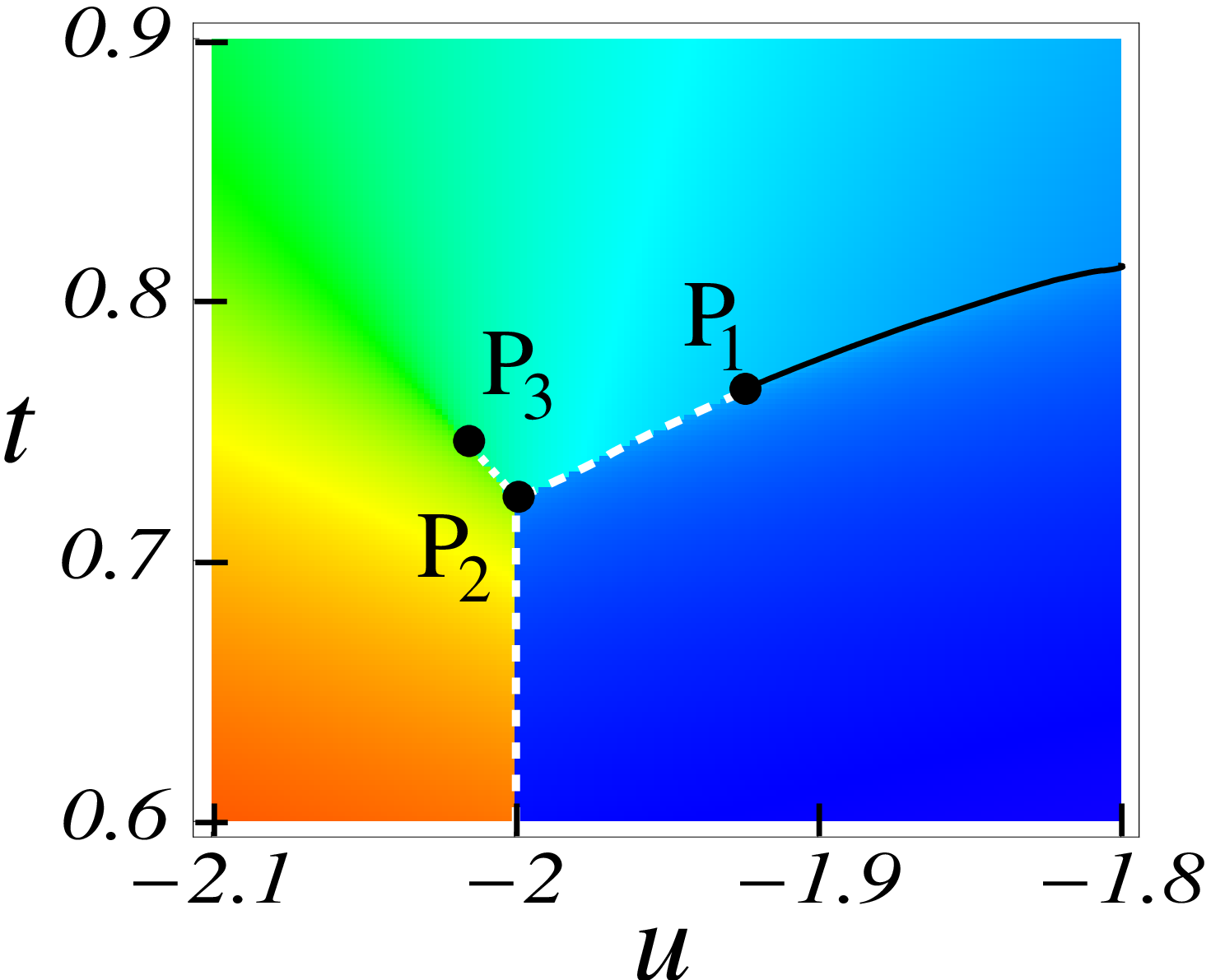}
\end{minipage}%
\begin{minipage}[c]{.10\textwidth}
\includegraphics[width = 0.8\textwidth,height = 3.\textwidth]{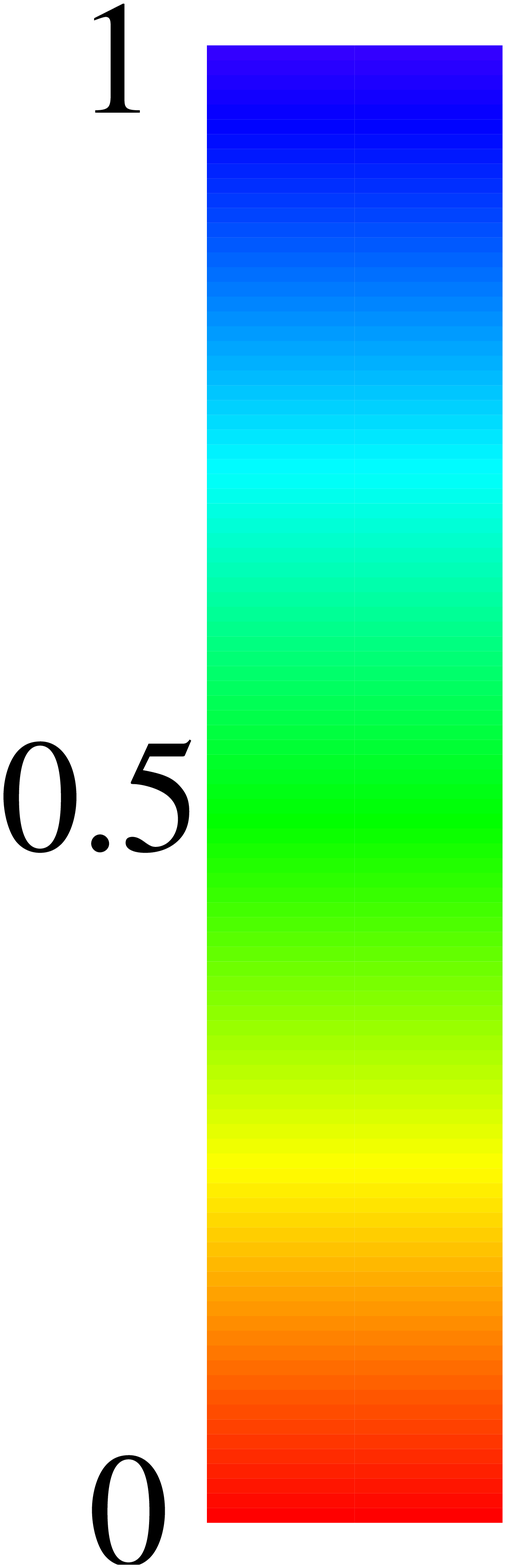}
\centerline{~\hfill}
\end{minipage}
\caption{
\label{fig2}
(a) Magnetization $M$ and (b) quadrupolar moment $Q$ for an inert 
substrate ($q = 0$) as functions of $t:= k_B T/(z J_0)$ and 
$u:= \mu /(z J_0)$ for $\kappa_0 = 3.0$. The color coding (shown at 
the right) is the same for both figures, and linearly interpolates 
between zero (red) and one (dark blue) via yellow, green, and light 
blue. $P_1$ is a tricritical point, $P_2$ is a quadruple point, and 
$P_3$ is a critical end point. Solid black lines are lines of 
critical points and indicate second-order phase transitions, while 
the dashed white lines are lines of triple points indicating 
first-order transitions.
}
\end{center}
\end{figure}

We now turn to a detailed discussion of the phase diagrams. 
The most complicated behavior occurs for intermediate values 
of $\kappa_0$, e.g., $\kappa_0 \simeq 3$ \cite{mean_field}. 
For $\kappa_0 = 3.0$, in Fig. \ref{fig2} we show (color coded) 
the order parameters $M = |n_A - n_B|$ and $Q = n_A + n_B$,  
as well as the lines corresponding to the various phase 
transitions composing the phase diagram.

At low temperatures, i.e. $t \leq t_{P_2} \simeq 0.72$ and low 
negative chemical potential, i.e., $u << u_{P_2} \simeq -2.0$, 
the substrate is covered by a very low density, two-dimensional 
mixed gas ($M = 0$, $Q \ll 1$), in agreement with 
(\textbf{ii}) above. Increasing the chemical potential $u$ at 
fixed temperature $t < t_{P_2}$, the system undergoes a first-order 
phase transition at $u = u_c(t) \simeq -2.0$ (the white 
dashed line located at $u_c(t) \simeq -2.0$) upon which the 
density of the monolayer increases abruptly to almost one 
[as indicated by the dark blue color in Fig. \ref{fig2}(b)] and 
the monolayer also (partially) demixes, i.e. $0 < M \leq Q$ 
[as indicated by the lighter blue or green color in 
Fig. \ref{fig2}(a)]. Thus in the region  $u > u_c(t)$ as anticipated 
in (\textbf{iv}) above the substrate is covered by a dense $A$-rich 
monolayer (for $H=\mu_A-\mu_B \to 0^+$; with equal probability, a 
dense $B$-rich monolayer forms for $H=\mu_A-\mu_B \to 0^-$). 
Therefore this first-order transition line, which is located at 
almost constant $\mu \simeq -2.0$ and extends from $P_2$ to $t \to 0$, 
is a line of triple points.

We focus now on the region $u > u_c(t)$. At constant $u$, upon 
approaching from below the line $P_2$-$P_1$, which continues to 
$u \to \infty$, the demixing is less and less 
pronounced. Increasing the temperature at fixed $u \geq u_{P_1}$,
upon crossing the line starting at $P_1$ (solid black line in 
Fig. \ref{fig2}) the dense $A$-rich (or $B$-rich) monolayer 
undergoes a second-order phase transition (both $M$ and $Q$ are 
changing continuously there; note in Fig. \ref{fig2}(a) the thin 
band of yellow color, which ends at $P_1$, corresponding to very 
small but non-zero values of $M$) such that at high temperatures 
the substrate is covered by a mixed dense monolayer. As discussed 
in (\textbf{i}) and (\textbf{iii}) above, this line of critical 
points stays below $t = 1$ for all values of $u$, and it approaches
 $t = 1$ for $u \to \infty$. Upon crossing the line segment 
$P_2$-$P_1$ (white dashed line) the transition from the dense 
$A$-rich (or $B$-rich) monolayer to the dense mixed monolayer is 
of first order with a jump in both coverage and composition, i.e., 
in both $Q$ and $M$ (using the color code, for $M$ this is indicated 
by the transition from red to green, without yellow in between, 
while for $Q$ by the direct transition from dark blue to light-blue 
and green). The line segment $P_2$-$P_1$ is also a line of triple 
points since there three phases coexist: dense $A$-rich, 
dense $B$-rich, and dense mixed monolayer, respectively. 

For $u < u_c(t)$, upon crossing (from below) the line $P_2 - P_3$ 
the system undergoes a first-order transition from a low density 
mixed monolayer to a higher density mixed monolayer: there is a 
jump in the coverage, i.e., $Q$ varies discontinuously [as 
indicated in Fig. \ref{fig2}(b) by the color change from light 
green to light blue (without dark green in between)], while $M$ 
remains zero. 
The jump in $Q$ upon crossing the line $P_2 - P_3$ decreases as 
the crossing point approaches $P_3$, and it becomes zero at $P_3$.
Thus $P_3$ is a critical point. Note that for temperatures $t$ such 
that $t_{P_2} < t < t_{P_3}$, e.g., $t = 0.73$, increasing $u$ at 
constant temperature from small negative values toward positive 
values will drive the state of the monolayer from the mixed gas 
phase towards the $A$-rich (or $B$-rich) dense phase via two 
consecutive first-order phase transitions, corresponding to 
crossing the line $P_2 - P_3$ (with a jump only in $Q$) 
followed by crossing the line $P_2 - P_1$ (with a jump in both $Q$ 
and $M$).

The point $P_1$ is a tricritical point (it belongs also to the 
critical lines of the $A$-rich dense phase $\to$ mixed gas and 
$B$-rich dense phase $\to$ mixed gas transitions). $P_2$ is a 
quadruple point, i.e., at $P_2$ four phases coexist: dense 
$A$-rich, dense $B$-rich, dense mixed, and dilute mixed monolayer, 
respectively 

Varying $\kappa_0$ will lead to topological changes only near the 
points $P_1$, $P_2$, $P_3$ as follows. For $\kappa_0 \ll 1$, the 
line $P_2$-$P_3$ and thus the point $P_2$ do not occur [there is 
no jump in the total coverage if the density is increased while 
keeping the monolayer mixed, i.e., in that region which is red 
in Fig. \ref{fig2}(a)], so that at the tricritical point $P_1$ 
the line of triple points connects directly with the one 
corresponding to the second-order phase transitions. With increasing 
$\kappa_0$, the line $P_2$-$P_3$ emerges, and for $\kappa_0 \lesssim 3$ 
the behavior is the same as the one at $\kappa_0 = 3$. For 
$\kappa_0 \gtrsim 3$ the only change is that the point $P_3$ is located 
at higher values of $t$ than $P_1$. With increasing $\kappa_0 > 3$ $P_1$ 
is shifting towards $P_2$ and eventually reaches the first-order triple 
line such that for large $\kappa_0$, e.g., $\kappa_0 \gtrsim 5$, $P_1$ 
(which at low and medium values of $\kappa_0$ is a tricritical point) 
merges with $P_2$ forming a critical end point.

\section{The case of a disordered substrate: $0 < q < 1$.}
\label{q_gen}
In the case of a disordered substrate, $0 < q < 1$, one has 
\begin{eqnarray}
J' = z J_0 - \frac{z k_B T}{2} \ln(1-q) \,, \nonumber\\
K' = z K_0 + \frac{z k_B T}{2} \ln(1-q) \,.
\end{eqnarray}
Using again $z J_0$ as the energy scale, the system is now 
characterized by the two variables $t$ and $u$ defined in 
Subsec. \ref{q_0}, a disorder parameter $\bar q$ defined as
\begin{equation}
\bar q = 1-(1-q)^z\,,\mathrm{~i.e.,~} 
z \ln(1-q)= \ln(1-\bar q)\,, 
\label{bar_q}
\end{equation}
and the parameter 
\begin{equation}
\kappa(t,\bar q) := K'/J' = 
\frac{2 \kappa_0 + t \ln(1-\bar q)}{2 - t \ln(1-\bar q)}\,.
\label{kappa_q}
\end{equation}
Here we indicated explicitely that now the ratio $\kappa$ 
depends both on the temperature and on the disorder parameter 
$\bar q$.[Note that $\kappa(t,0) = \kappa(0,\bar q) = \kappa_0$.] 
Therefore, we shall discuss the phase diagrams, as well as the 
behavior of the order parameters $M$ and $Q$, in the $u - t$ 
plane for given values of $\kappa_0$ and $\bar q$.

Eq.(\ref{kappa_q}) implies that for any given $0 < \bar q < 1$ 
and $\kappa_0 > 0$, the ratio $\kappa(t,\bar q)$ becomes negative 
at high enough temperatures, i.e., $\kappa(t > t_{tr},\bar q) < 0$, 
where the threshold temperature $t_{tr}$ is given by
\begin{equation}
t_{tr}(\bar q;\kappa_0) = 2 \kappa_0/|\ln(1-\bar q)|\,;
\label{temp_tr}
\end{equation}
note that for a given $\kappa_0$, i.e., for a given mixture, 
$t_{tr}$ is a decreasing function of $\bar q$. As already 
mentioned, for sufficiently negative values of $\kappa$ the 
system splits into ordered sub-lattices \cite{stagerred}, 
which generally leads to a significant decrease in the 
yield of the catalytic reaction. Using $t_{tr}$ as a measure 
of this tendency, Eq.(\ref{temp_tr}) implies that it is desirable 
to run the reaction at low enough temperatures in order to 
maintain a mixed monolayer, and that this range of temperatures 
decreases with an increasing mean density of catalytic bonds.

\begin{figure}[!htb]
\begin{center}
\centerline{\hspace*{.2\textwidth}(a) 
[M]\hspace*{.37\textwidth}(b) [Q]\hfill}
\begin{minipage}[c]{.45\textwidth}
\includegraphics[width = 0.9\textwidth,height=0.9\textwidth]{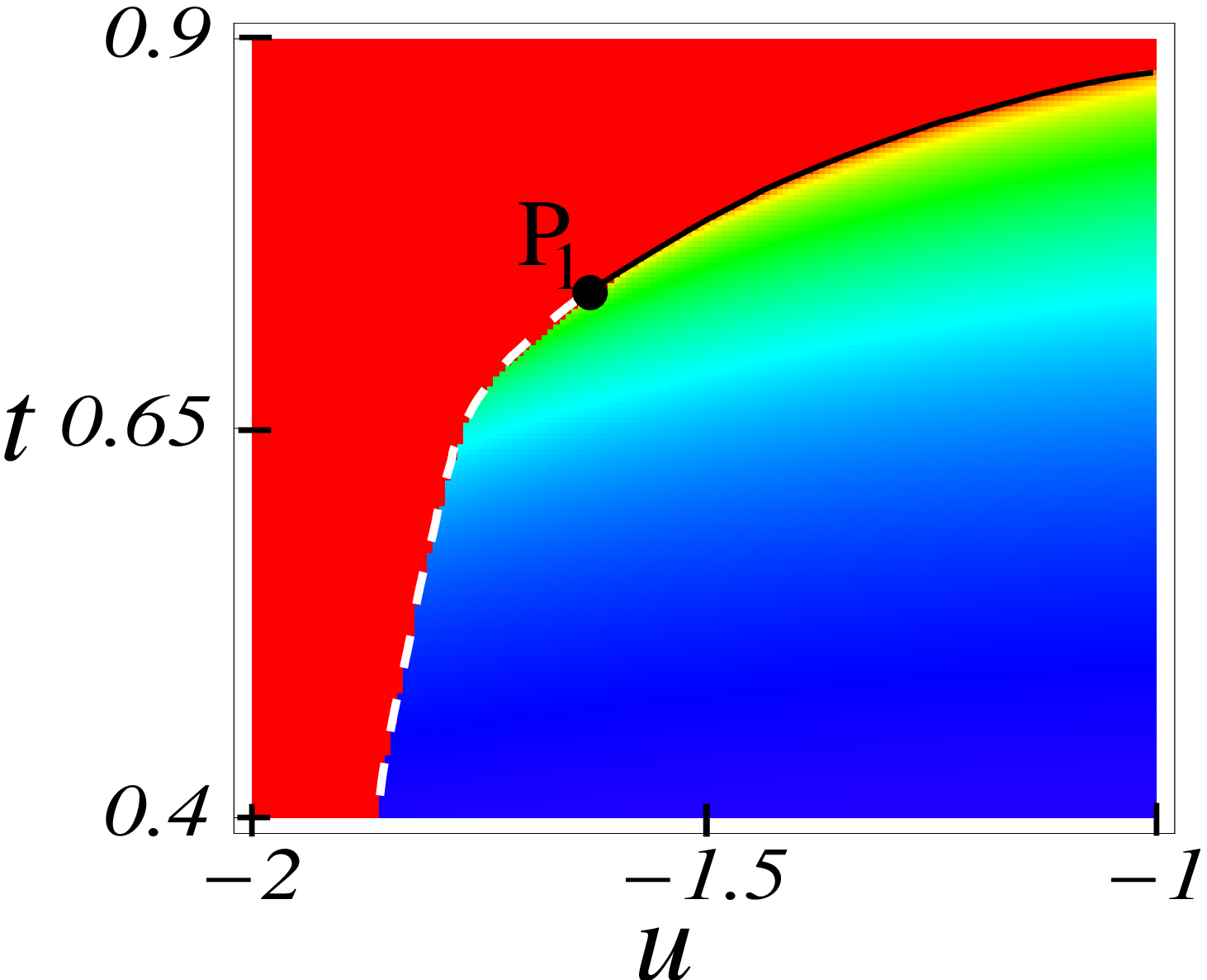}
\end{minipage}%
\begin{minipage}[c]{.45\textwidth}
\includegraphics[width = 0.9\textwidth,height=0.9\textwidth]{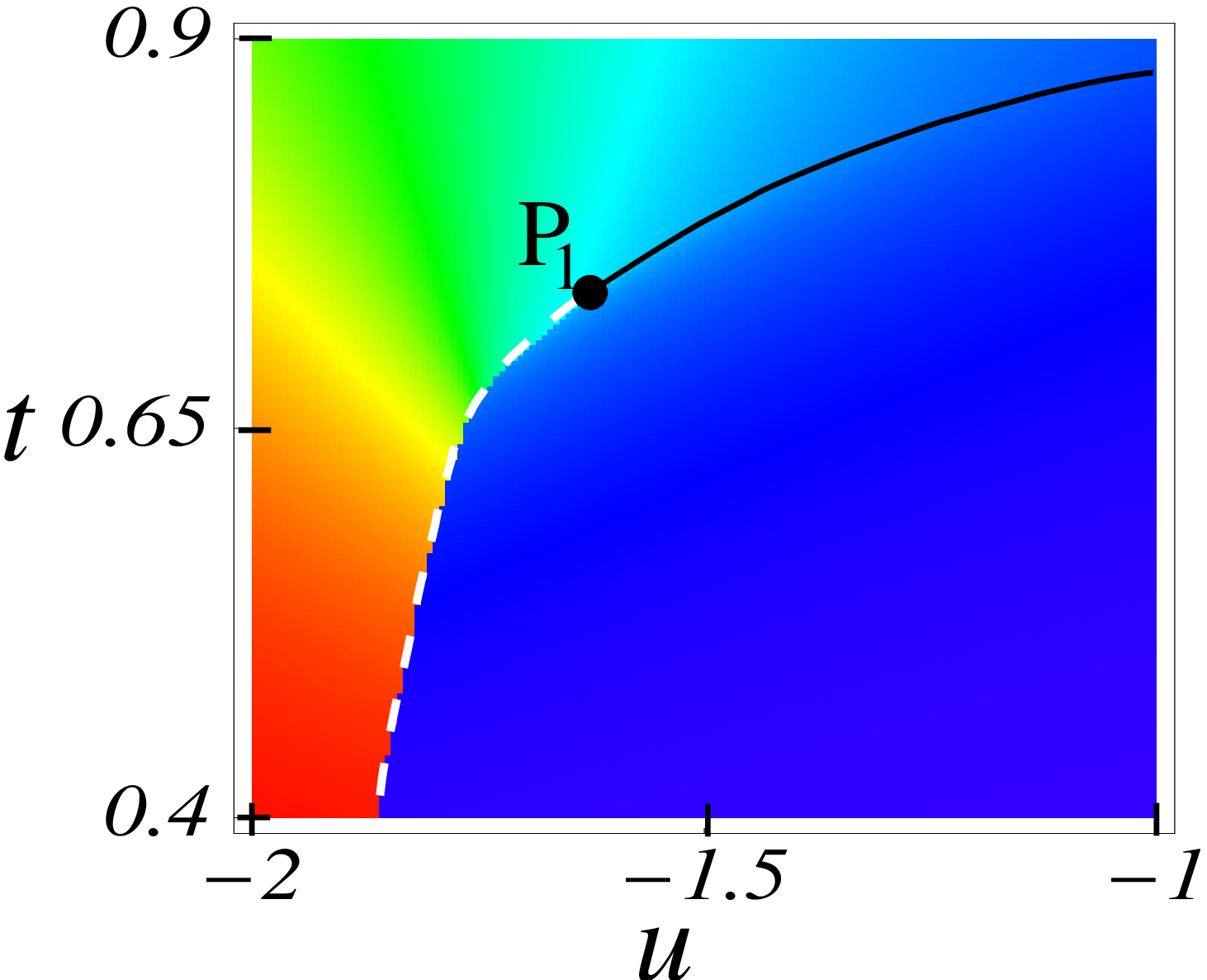}
\end{minipage}%
\begin{minipage}[c]{.10\textwidth}
\includegraphics[width = 0.8\textwidth,height = 3.\textwidth]{fig2c.eps}
\centerline{~\hfill}
\end{minipage}
\caption{
\label{fig3}
(a) Magnetization $M$ and (b) quadrupolar moment $Q$ as 
functions of $t:= k_B T/(z J_0)$ and $u:= \mu /(z J_0)$ 
for $\kappa_0 = 3.0$ and $\bar q = 0.3$. 
The color coding (shown at the right) is the same for both figures, 
and linearly interpolates between zero (red) and one (dark blue) 
via yellow, green, and light blue. Solid black 
lines are lines of critical points and indicate second-order 
phase transitions, while the dashed white lines are lines of 
triple points indicating first-order transitions.
}
\end{center}
\end{figure}

\begin{figure}[!htb]
\begin{center}
\centerline{\hspace*{.2\textwidth}(a) 
[M]\hspace*{.37\textwidth}(b) [Q]\hfill}
\begin{minipage}[c]{.45\textwidth}
\includegraphics[width = 0.9\textwidth,height=0.9\textwidth]{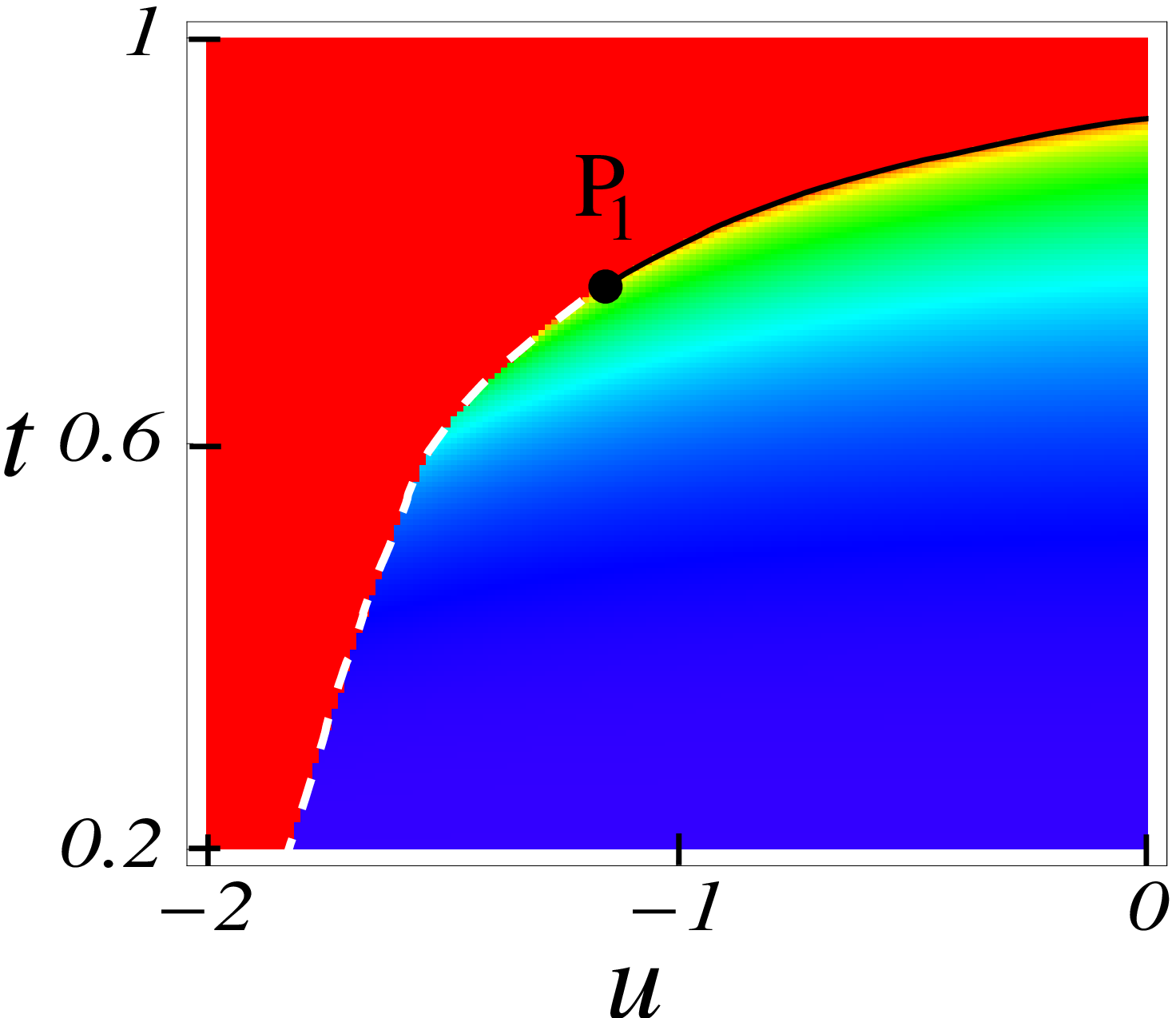}
\end{minipage}%
\begin{minipage}[c]{.45\textwidth}
\includegraphics[width = 0.9\textwidth,height=0.9\textwidth]{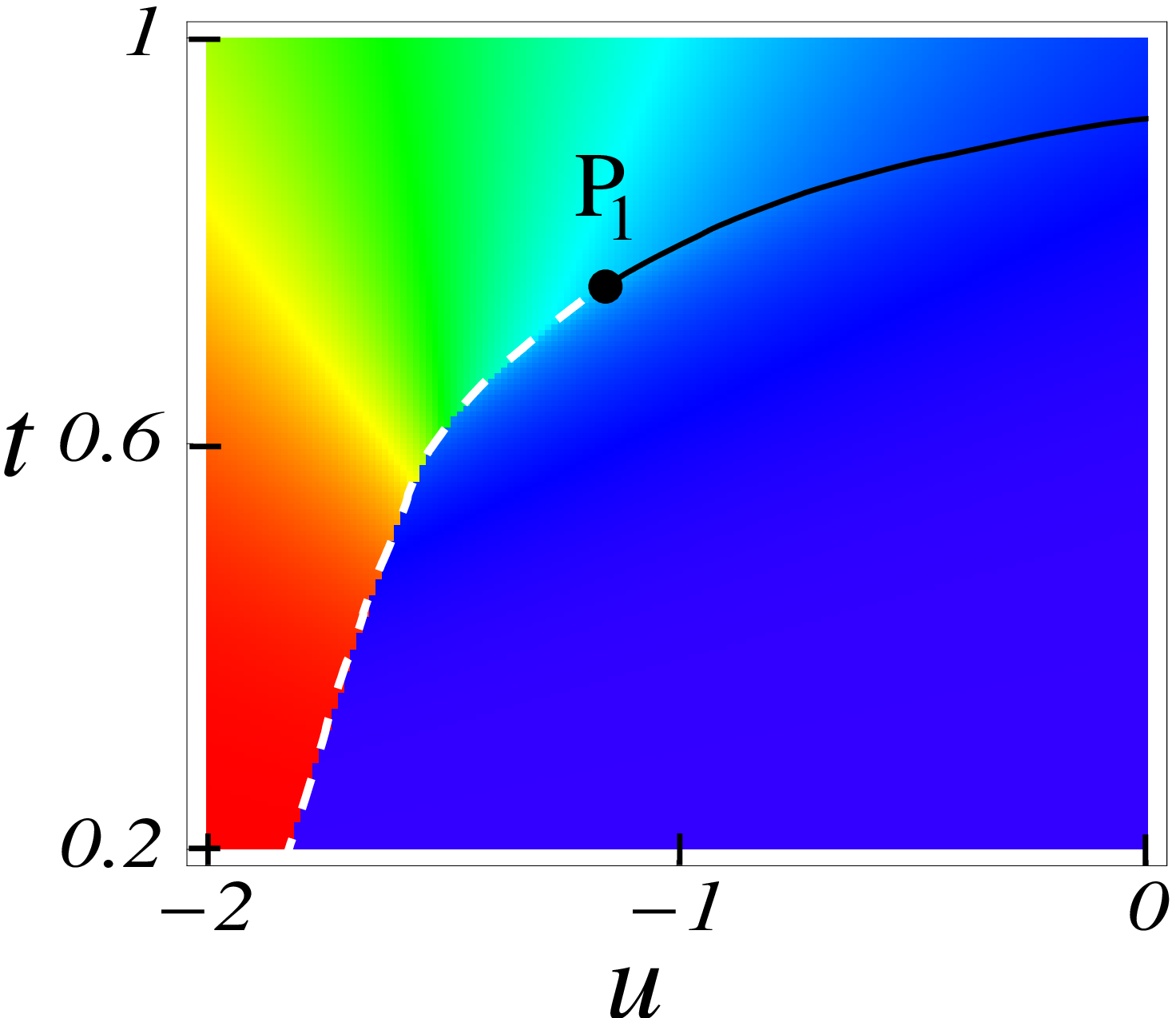}
\end{minipage}%
\begin{minipage}[c]{.10\textwidth}
\includegraphics[width = 0.8\textwidth,height = 3.\textwidth]{fig2c.eps}
\centerline{~\hfill}
\end{minipage}
\caption{
\label{fig4}
Same as Fig. \ref{fig3} for $\bar q = 0.6$.
}
\end{center}
\end{figure}

At constant temperature, and for fixed values of $\bar q$ and 
of $\kappa_0$, the parameter $\kappa(t,\bar q)$ is constant; 
for two temperatures $t_1$ and $t_2 > t_1$, it satisfies 
$\kappa(t_2 > t_1,\bar q) < \kappa(t_1,\bar q)$ (in particular, 
one has $\kappa(t,\bar q) \leq \kappa_0$). Thus, for any 
chosen temperature $\bar t \leq t_{tr}(\bar q;\kappa_0)$ the 
isotherms $M(\bar t,u)$ and $Q(\bar t,u)$ can be simply read 
as the ones corresponding to an inert substrate, as in 
Subsec. \ref{q_0}, at the same temperature $\bar t$ but for a 
binary mixture with 
$\bar \kappa_0 = \kappa(\bar t,\bar q) < \kappa_0$. Since 
larger values of $\bar t$ correspond to smaller values of 
$\bar \kappa_0$, the phase diagram is expected to be more 
similar to one at low $\kappa_0$ on an inert substrate, and 
thus only the first-order mixed gas $\to$ $A$-rich ($B$-rich) 
dense monolayer transition and the second-order transition 
lines joining at a tricritical point are generally present. 
The tricritical point $P_1$ shifts towards increasing values of 
$u$ with increasing $\bar q$, and the first-order transition line 
is a smooth curve (rather than consisting of two segments 
$P_2-P_1$ and $P_2 \to$ zero temperature), which runs from 
$u \simeq -2.0$ at low temperature (corresponding to the location 
in the case $\kappa_0$) to $P_1$.
These qualitative features can be seen in Figs. \ref{fig3} 
and \ref{fig4} where we show results corresponding to 
$\kappa_0 = 3.0$ for $\bar q = 0.3$ and $\bar q = 0.6$, 
respectively, which allows one a direct comparison with the 
phase diagram on the inert substrate. Note that in these cases the 
threshold temperatures [Eq.(\ref{temp_tr})] for splitting into 
ordered sub-lattices are very high [$t_{tr}(0.3;0.3) \simeq 16.8$, 
$t_{tr}(0.6;0.3) \simeq 6.54$], thus the present version of the 
mean-field analysis is justified in the range $t \leq 1$ we are 
interested in.

For an efficient catalytic reaction, the system would have to be 
operated such that the substrate is covered by a mixed, not too 
dilute monolayer, i.e., not too low positive chemical potential and 
temperatures above the second-order transition line, i.e., 
$t \simeq 1.0$, but below the threshold temperature 
$t_{tr}(\bar q;\kappa_0)$ for which the splitting into ordered 
sublattices may occur. For a given binary mixture, i.e., given 
$\kappa_0$, the requirement $t_{tr} < 1$ implies an upper 
bound $\bar q_{op}$ on the mean density of catalytic bonds, i.e.,
\begin{equation}
t_{tr} < 1  \Rightarrow \bar q < \bar q_{op}=1-\exp(-2\kappa_0)\,.
\end{equation}
This is somewhat surprising: a substrate which is only partially 
decorated by the catalyst, i.e., $\bar q < \bar q_{op} <  1$, would be 
the optimal choice because it avoids transitions into a passive 
state (poisoned substrate). For binary mixtures with small values 
of $\kappa_0$ the upper bound given above implies a drastic 
constraint (e.g., at $\kappa_0 = 0.5$, $\bar q_{op} \simeq 0.63$), 
while for binary mixtures with large values of $\kappa_0$, the above 
constraint is basically irrelevant 
(e.g., at $\kappa_0 = 3.0$, $\bar q_{op} \simeq 0.998$).

\section{Summary}
\label{summary}
Within a mean-field approach we have studied the equilibrium 
properties of a model for a binary mixture of catalytically-reactive 
monomers adsorbed on a two-dimensional substrate, decorated by 
randomly placed catalytic bonds of mean density $q$. Our analysis 
here has been focused on annealed disorder in the placement 
of the catalytic bonds and on the symmetric case in 
which the chemical potentials $\mu_A$ and $\mu_B$, as well as the 
interactions $J_A$ and $J_B$ of the two species, are equal.

We have shown that in the general case $0 < q < 1$ the mean-field 
phase diagram and the behavior of the composition $n_A-n_B$ and of 
the total coverage $n_A+n_B$ can be extracted from the corresponding
results on an inert substrate ($q=0$). We have determined certain 
restrictions on the temperature, at which the system is operated, 
as well as a somewhat surprising upper bound $q_{op}$ of the density 
of catalytic bonds, which have to be obeyed in order to maintain the 
monolayer in a mixed and thus active state. Even in this highly 
symmetric case studied here, the phase diagrams are rather rich, 
containing, e.g., lines of second-order phase transitions and 
tricritical points. We have pointed out the likely occurrence of 
staggered phases at intermediate temperatures.

Finally, we note that in the generic case in which $\mu_A \neq \mu_B$ 
and  $J_A \neq J_B$, the behavior is expected to be even richer (see, 
e.g., Ref. \cite{pd_gen_S1} for a detailed discussion of the 
phase diagrams for the general spin $S =1$ model). These aspects, as 
well as the issue of staggered phases or that of \textit{quenched} disorder, are left for future work.

\end{document}